\documentclass[showpacs,preprintnumbers,
amsmath,amssymb,aps,prd]{revtex4}
\usepackage{amsfonts}
\usepackage{amsfonts,graphics,epsfig,subfigure}

\begin{document}

\title{Effects of power-law Maxwell field on the critical phenomena of higher dimensional dilaton black holes}
\author{Jie-Xiong Mo \footnote{Corresponding author: mojiexiong@gmail.com}, Gu-Qiang Li \footnote{zsgqli@hotmail.com}, Xiao-Bao Xu \footnote{xbxu789@163.com}}

 \affiliation{Institute of Theoretical Physics, Lingnan Normal University, Zhanjiang, 524048, Guangdong, China}

\begin{abstract}
The effects of power-law Maxwell field on the critical phenomena of higher-dimensional dilaton black holes are probed in detail. We successfully derive the analytic solutions of critical point and carry out some check to ensure that these critical quantities are positive. It is shown that the constraint on the parameter $\alpha$ describing the strength of coupling of the electromagnetic field and the scalar field turns out to be $0<\alpha^2<1$, which is more tighter than that in the non-extended phase space. It is also shown that these critical quantities and the ratio $P_cv_c/T_c$ are affected by the power-law Maxwell field. Moreover, critical exponents are found to coincide with those of other AdS black holes, showing the powerful influence of mean field theory.

\end{abstract}

\keywords{dilaton black holes\;  power-law Maxwell field\; critical phenomena}

 \pacs{04.70.Dy, 04.70.-s} \maketitle

\section{Introduction}
     Dilaton field is a scalar field that appears in the low energy limit of string theory. One or more Liouville-type potentials, resulted by the breaking of spacetime supersymmetry in ten dimensions, are contained in the action of dilaton gravity. The dilaton field exerts such important influences on casual structure and thermodynamics of black holes that both its black hole solutions and thermodynamic properties have attracted extensive attention of researchers \cite{Gibbons1}- \cite{Sheykhi11}. In an intersting paper \cite{Sheykhi13}, the authors considered a linearly charged dilatonic black holes and studied their thermodynamical behavior in the context of phase transition and thermodynamic geometry. They employed a novel thermodynamical geometry metric and showed that it worked perfectly especially when dealing with the case that dilaton parameter is viewed as extensive parameter.

     Recently, Ref.  \cite{Sheykhi8} constructed a novel class of higher dimensional topological black hole solutions in Einstein-dilaton gravity coupled to power-law Maxwell filed. These solutions are of great physical significance. Because in higher dimensions standard Maxwell Lagrangian is not conformally invariant while the Lagrangian with a power-law Maxwell filed is conformally invariant if the parameter $p$ describing the nonlinearity of the electromagnetic field satisfies the relation $p=(n+1)/4$. The relevant physical quantities, such as the mass, temperature, charge, electric potential were derived~\cite{Sheykhi8}. Thermal stability of these black holes was also thoroughly investigated~\cite{Sheykhi8}. In this paper, we would like to generalize the thermodynamics to the extended phase space, where we regard the cosmological constant as the thermodynamic pressure.

     It is of vital importance to study the thermodynamics of black holes in the extended phase space. Firstly, only in the extended phase space can the first law of thermodynamics of AdS black holes matches the Smarr relation. Secondly, the variation of physics constants can be admitted in more fundamental theories. Thirdly, the mass of the black hole should be interpreted as enthalpy rather than internal energy. This treatment will help understand the relations between critical behaviours of AdS black holes and those of ordinary thermodynamic systems. Recently, $P-V$ criticality of charged AdS black holes was investigated in the extended space~\cite{Kubiznak} where the analogy between charged AdS black holes and Van der Waals liquid-gas systems was enhanced. Furthermore, reentrant phase transitions of black holes were found to be reminiscent of multicomponent liquids ~\cite{Gunasekaran,Altamirano1,Frassino,Hennigar,Shaowen1,Altamirano3} while Small/intermediate/large black hole phase transitions reminiscent of solid/liquid/gas phase transition were reported~\cite{Frassino,Altamirano2,Shaowen1}. The research on the extend phase space thermodynamics of black holes has been fruitful in the past few years~\cite{Kastor}-\cite{Lee} . For nice reviews, see Ref.~\cite{Altamirano3,Dolan4}.

    Among the literatures, there have been efforts on the effect of dilaton field on the extended phase space thermodynamics~\cite{Sheykhi1,Zhao,zhao2} and efforts on the effect of power-law Maxwell field~\cite{Sheykhi8,Sheykhi13,Sheykhi12,Hendi1}. However, to the best of our knowledge, the combined effects of both the power-law Maxwell field and the dilaton field on the Van der Waals like phase transition have not been reported yet. It is expected that some unique characteristics will be disclosed, motivating us to carry out the research in this paper. On the other hand, obtaining the analytic expressions of critical quantities is a nontrivial task and deserves to be investigated since the situation will be quite complicated when both the power-law Maxwell field and dilaton field are taken into consideration.

    The organization of this paper is as follows. Relevant physical quantities of topological dilaton black holes with power-law Maxwell field will be reviewed briefly in Sec. \ref{Sec2}. Then we will investigate their extended phase space thermodynamics in Sec. \ref{Sec3}. Critical exponents will be calculated in Sec. \ref{Sec4}. In Sec. \ref{Sec5}, we will present a brief conclusion.

\section{Review of topological dilaton black holes with power-law Maxwell field}
\label{Sec2}
The $(n+1)$-dimensional action where gravity is coupled to a dilaton field and power-law Maxwell filed reads ~\cite{Sheykhi8}
\begin{equation}
S=\frac{1}{16\pi}\int d^{n+1}x \sqrt{-g}\left\{\mathfrak{R}-\frac{4}{n-1}(\nabla \Phi)^2-V(\Phi)+(-e^{-4\alpha \Phi/(n-1)}F)^p\right\},\label{1}\\
\end{equation}
where $\Phi$ is the dilaton field whose potential is denoted as $V(\Phi)$. $F=F_{\mu\nu}F^{\mu\nu}$, where $F_{\mu\nu}=\partial_{\mu}A_{\nu}-\partial_{\nu}A_{\mu}$ is the electromagnetic field tensor. $\alpha$ describes the strength of coupling of the electromagnetic field and scalar field while $p$ describes the nonlinearity of the electromagnetic field.

It has been demonstrated in Ref.~\cite{Sheykhi8} that one can find topological black hole solutions if the above dilaton potential consists of three Liouville-type potentials as follow
\begin{equation}
V(\Phi)=2\Lambda_1 e^{2\zeta_1\Phi}+2\Lambda_2 e^{2\zeta_2\Phi}+2\Lambda e^{2\zeta_3\Phi}.\label{2}\\
\end{equation}
Note that $\Lambda_1$ and $\Lambda_2$ are fixed by Eq.(\ref{5}) while $\Lambda$ remains a free parameter playing the role of the cosmological constant with the relation $\Lambda=-n(n-1)/2l^2$ holds~\cite{Sheykhi8}.

The corresponding solution has been derived as~\cite{Sheykhi8}
\begin{eqnarray}
ds^2&=&-f(r)dt^2+\frac{1}{f(r)}dr^2+r^2R^2(r)h_{ij}dx^idx^j,\label{3}\\
F_{tr}&=&\frac{qe^{\frac{4\alpha p\Phi(r)}{(n-1)(2p-1)}}}{(rR)^{\frac{n-1}{2p-1}}},\label{997}\\
\Phi(r)&=&\frac{(n-1)\alpha}{2(\alpha^2+1)}\ln \left(\frac{b}{r}\right),\label{998}
\end{eqnarray}
where
\begin{eqnarray}
f(r)&=&\frac{k(n-2)(1+\alpha^2)^2r^{2\gamma}}{(1-\alpha^2)(\alpha^2+n-2)b^{2\gamma}}-\frac{m}{r^{(n-1)(1-\gamma)-1}}-\frac{2\Lambda b^{2\gamma}(1+\alpha^2)^2r^{2(1-\gamma)}}{(n-1)(n-\alpha^2)}
\nonumber
\\
&\,&+\frac{2^pp(1+\alpha^2)^2(2p-1)^2b^{-\frac{2(n-2)p\gamma}{2p-1}}q^{2p}}{\Pi(n+\alpha^2-2p)r^{-\frac{2[(n-3)p+1]-2p(n-2)\gamma}{2p-1}}},\label{4}\\
R(r)&=&e^{2\alpha \Phi(r)/(n-1)}.\label{999}
\end{eqnarray}

Note that the solution exists for conformally invariant source $p=(n+1)/4$~\cite{Sheykhi8}. And it reduces to the Einstein-Maxwell-dilaton topological black hole when $p=1$. $h_{ij}dx^idx^j$ is the line element of an $(n-1)$-dimensional hypersurface with constant scalar curvature $(n-1)(n-2)k$.  $k$ is a constant characterizing the hypersurface. $k$ can be taken as $-1,\,0,\,1$, corresponding to hyperbolic, flat and spherical constant curvature hypersurface respectively. $b$ is an arbitrary positive constant while $q$ is an integration constant related to the electric charge of the black hole. $m$ is a constant related to the mass of the black hole by Eq. (\ref{9}). Relations of other constants are listed as follows~\cite{Sheykhi8}
\begin{eqnarray}
\gamma&=&\frac{\alpha^2}{\alpha^2+1},\;\;\;\Pi=\alpha^2+(n-1-\alpha^2)p,\nonumber
\\
\zeta_1&=&\frac{2}{(n-1)\alpha},\;\;\;\zeta_2=\frac{2p(n-1+\alpha^2)}{(n-1)(2p-1)\alpha},\;\;\;\zeta_3=\frac{2\alpha}{n-1},\nonumber
\\
\Lambda_1&=&\frac{k(n-1)(n-2)\alpha^2}{2b^2(\alpha^2-1)},\;\;\;\Lambda_2=\frac{2^{p-1}(2p-1)(p-1)\alpha^2q^{2p}}{\Pi b^{\frac{2(n-1)p}{2p-1}}}.\label{5}
\end{eqnarray}

Considering the fact that the electric potential $A_t$ should be finite at infinity and the fact that the term including
 $m$ should vanish at spacial infinity, the restrictions on $p$ and $\alpha$ has been derived as follow~\cite{Sheykhi8}
 \begin{eqnarray}
&\,&For \;\frac{1}{2}<p<\frac{n}{2},\;\;\;0\leq\alpha^2<n-2,\nonumber
\\
&\,&For \;\frac{n}{2}<p<n-1,\;\;\;2p-n<\alpha^2<n-2.\label{6}
\end{eqnarray}

Solving the equation $f(r_+) =0$ for the positive real root $r_+$, the expression of $m$ can be obtained as
\begin{eqnarray}
m&=&\frac{k(n-2)b^{-2\gamma}r_+^{\frac{\alpha^2+n-2}{\alpha^2+1}}}{(2\gamma-1)(\gamma-1)(\alpha^2+n-2)}-\frac{2b^{2\gamma}\Lambda r_+^{\frac{n-\alpha^2}{\alpha^2+1}}}{(n-1)(\gamma-1)^2(n-\alpha^2)}
\nonumber
\\
&\,&+\frac{2^pp(2p-1)^2b^{-\frac{2(n-2)p\gamma}{2p-1}}q^{2p}r_+^{-\frac{\alpha^2-2p+n}{(2p-1)(\alpha^2+1)}}}{\Pi(\gamma-1)^2(\alpha^2-2p+n)}.\label{7}
\end{eqnarray}

The Hawking temperature can be calculated as
\begin{equation}
T=\frac{f'(r_+)}{4\pi}=\frac{1+\alpha^2}{4\pi}\Big[\frac{k(n-2)}{b^{2\gamma}(1-\alpha^2)r_+^{1-2\gamma}}-\frac{2\Lambda b^{2\gamma} r_+^{1-2\gamma}}{n-1}-\frac{2^pp(2p-1)b^{\frac{-2(n-2)p\gamma}{2p-1}}}{\Pi r_+^{\frac{2p(n-2)(1-\gamma)+1}{2p-1}}q^{-2p}}\Big].\label{8}
\end{equation}

The mass, the entropy, the charge and the electric potential has been obtained as~\cite{Sheykhi8}
\begin{eqnarray}
M&=&\frac{\omega_{n-1}b^{(n-1)\gamma}(n-1)m}{16\pi(\alpha^2+1)},\label{9} \\
S&=&\frac{\omega_{n-1}b^{(n-1)\gamma}r_+^{(n-1)(1-\gamma)}}{4},\label{10} \\
Q&=&\frac{\omega_{n-1}2^{p-1}q^{2p-1}}{4\pi},\label{11} \\
U&=&\frac{(n-1)p^2qb^{\frac{(2p-n+1)\gamma}{2p-1}}}{\Pi\Upsilon r_+^\Upsilon},\label{12}
\end{eqnarray}
where $\Upsilon=\frac{n-2p+\alpha^2}{(2p-1)(1+\alpha^2)}$. Note that the expressions of mass, entropy and charge in Ref.~\cite{Sheykhi8} correspond to the unit volume. We have recovered in the above expressions the factor $\omega_{n-1}$, which denotes the volume of the unit $(n-1)$-sphere.

\section{Extended phase space thermodynamics of topological dilaton black holes with power-law Maxwell field}
\label{Sec3}
With the physical quantities presented above, one can carry out some investigation in the extended phase space, where the cosmological constant and its conjugate quantity can be interpreted as thermodynamic pressure and volume respectively as follow
\begin{eqnarray}
P&=&-\frac{\Lambda}{8\pi},\label{13}
\\
V&=&\left(\frac{\partial M}{\partial P}\right)_{S,Q}.\label{14}
\end{eqnarray}
With Eqs. (\ref{7}), (\ref{9}), (\ref{13}) and (\ref{14}), the thermodynamic volume can be calculated as
\begin{equation}
V=\frac{\omega_{n-1}b^{(1+n)\gamma}r_+^{\frac{n-\alpha^2}{1+\alpha^2}}}{(n-\alpha^2)(1+\alpha^2)(\gamma-1)^2},\label{15}
\end{equation}
Note that Eq. (\ref{15}) is exactly the same as the thermodynamic volume of Einstein-Maxwell-dilaton black holes \cite{Zhao}, implying that power-law Maxwell field does not affect the thermodynamic volume. When $n=3,\alpha=\gamma=0$, Eq. (\ref{15}) reproduces the result $V=\frac{4}{3}\pi r_+^3$ of four-dimensional RN-AdS black hole.

The first law of black hole thermodynamics in the extended phase space can be written as
\begin{equation}
dM=TdS+U dQ+VdP.\label{16}
\end{equation}%

With Eqs. (\ref{8}) and (\ref{13}), one can derive the equation of state as
\begin{equation}
P=\frac{(n-1)r_+^{2\gamma-1}}{4(1+\alpha^2)b^{2\gamma}}\left[T-\frac{k(n-2)r_+^{2\gamma-1}(1+\alpha^2)}{4\pi (1-\alpha^2)b^{2\gamma}}+\frac{2^{p-2}p(2p-1)q^{2p}(1+\alpha^2)}{\pi\Pi b^{\frac{2(n-2)p\gamma}{2p-1}}r_+^{\frac{1+2(n-2)p(1-\gamma)}{2p-1}}}\right].\label{19}
\end{equation}

One can identify the specific volume as
\begin{equation}
v=\frac{4(1+\alpha^2)b^{2\gamma}}{(n-1)r_+^{2\gamma-1}}.\label{20}
\end{equation}

Then Eq. (\ref{19}) can be rewritten as
\begin{equation}
P=\frac{T}{v}-\frac{k(n-2)(1+\alpha^2)^2}{(n-1)\pi (1-\alpha^2)v^2}+\frac{2^{p-4}p(2p-1)q^{2p}(n-1)}{\pi [\frac{4(1+\alpha^2)}{(n-1)v}]^{\frac{2\Pi}{(2p-1)(\alpha^2-1)}}b^{\frac{2\gamma (np-2p+1)}{(2p-1)(2\gamma-1)}}\Pi}.\label{21}
\end{equation}

The critical point can be derived through the following conditions
\begin{eqnarray}
\left.\frac{\partial P}{\partial v}\right|_{T=T_c}&=&0,\label{22}\\
\left.\frac{\partial ^2P}{\partial v^2}\right|_{T=T_c}&=&0.\label{23}
\end{eqnarray}

However, Eq. (\ref{21}) is so complicated that the above equations can not be solved easily without some simplification. We introduce the notations as follow.
\begin{eqnarray}
A&=&\frac{k(n-2)(1+\alpha^2)^2}{(n-1)\pi (1-\alpha^2)},\label{24}\\
B&=&\frac{2\Pi}{(2p-1)(\alpha^2-1)},\label{25}\\
C&=&\frac{2^{p-4}p(2p-1)q^{2p}(n-1)}{\pi [\frac{4(1+\alpha^2)}{(n-1)}]^B b^{\frac{2\gamma (np-2p+1)}{(2p-1)(2\gamma-1)}}\Pi}.\label{26}
\end{eqnarray}
Then Eq. (\ref{21}) can be reorganized as
\begin{equation}
P=\frac{T}{v}-\frac{A}{v^2}+\frac{C}{v^{-B}}.\label{27}
\end{equation}

Substituting Eq. (\ref{27}) into Eqs. (\ref{22}) and (\ref{23}), one can analytically solve the equations and obtain
\begin{eqnarray}
v_c^{2+B}&=&\frac{2A}{(B+B^2)C},\label{28}\\
T_c&=&\frac{2A(2+B)}{(1+B)v_c},\label{29}\\
P_c&=&\frac{A(2+B)}{Bv_c^2}.\label{30}
\end{eqnarray}

To ensure that the above quantities are all positive, the terms $A$, $B$ and $C$ should satisfy
\begin{equation}
A>0, B\in(-\infty,-2)\cup(0,+\infty), C>0 \;\;  Or \; A<0, -2<B<-1, C<0.\label{31}
\end{equation}
Solving the above inequalities, one can derive the constraint on the parameters as $0<\alpha^2<1$.

Note that Eqs. (\ref{28})-(\ref{30}) are obtained by analytically solving the critical condition equations related to the equation of state (\ref{27}). In mathematical sense, it is the general solution. However, we have to remind the readers of the following two points who want to use these results directly in their future research. Firstly, they should strictly reorganize their equation of state into the form of Eq. (\ref{27}). Secondly, check with Eq. (\ref{31}) and also other specific physical conditions carefully to ensure whether the critical quantities make sense physically.

From Eqs. (\ref{25}) and (\ref{26}), one can see clearly that the values of $B$ and $C$ are influenced by the factor $p$, showing the effects of power-law Maxwell field. Consequently, the critical quantities are also influenced by the power-law Maxwell field. It can be proved that the above results reduce to those of charged dilaton AdS black holes reported in literature when $p=1$~\cite{Zhao}. Especially when $k=1,n=3,p=1,\alpha=\gamma=0$, one can get
\begin{eqnarray}
A&=&\frac{1}{2\pi},\;B=-4,\;C=\frac{2q^2}{\pi},\nonumber \\
v_c&=&2\sqrt{6}q,\;T_c=\frac{1}{3\sqrt{6}\pi q},\;P_c=\frac{1}{96\pi q^2},\label{32}
\end{eqnarray}
which exactly match the critical quantities for RN-AdS black holes~\cite{Kubiznak}.

Utilizing Eqs. (\ref{29}) and (\ref{30}), the ratio $P_cv_c/T_c$ can be calculated as
\begin{equation}
\frac{P_cv_c}{T_c}=\frac{1+B}{2B}=\frac{1+2np-4p+\alpha^2}{4\alpha^2+4np-4p-4\alpha^2p},\label{33}
\end{equation}%
which reduces to $3/8$ when $n=3,p=1,\alpha=0$ and recovers the result of four-dimensional RN-AdS black holes~\cite{Kubiznak}. The ratio gains corrections related to $p$ due to the effect of power-law Maxwell field.

The Gibbs free energy can be derived as
\begin{equation}
G=H-TS=M-TS.\label{34}
\end{equation}%
Note that we have interpreted the mass of black holes as enthalpy in the extended phase space.

Utilizing Eqs. (\ref{7}), (\ref{8}), (\ref{9}) and (\ref{10}), one can obtain
\begin{eqnarray}
G&=&(1+\alpha^2)\Omega_{n-2}\Big[\frac{k(n-2)b^{(n-3)\gamma}r_+^{3\gamma-n\gamma-2-n}}{16\pi(n+\alpha^2-2)}+\frac{b^{(n+1)\gamma}Pr_+^{n-\gamma-n\gamma}(1-\alpha^2)}{(n-1)(n-\alpha^2)}
\nonumber
\\
&\,&+\frac{b^{\frac{2p\gamma-n\gamma+\gamma}{2p-1}}2^pp(2p-1)q^{2p}(2pn-4p+1+\alpha^2)r_+^{\frac{2p-2p\gamma-n-\gamma+n\gamma}{2p-1}}}{16\pi (n-2p+\alpha^2) \Pi }\Big].\label{35}
\end{eqnarray}%

As a specific example, the case $p=2,n=7,k=1$ will be discussed thoroughly. The equation of state and Gibbs free energy for this case can be derived through Eqs. (\ref{21}) and (\ref{35}) as
\begin{eqnarray}
P&=&\frac{T}{v}-\frac{5(1+\alpha^2)^2}{6\pi (1-\alpha^2)v^2}+\frac{9q^{4}}{\pi [\frac{2(1+\alpha^2)}{3v}]^{\frac{2(12-\alpha^2)}{3\alpha^2-3}}b^{\frac{22\gamma }{3(2\gamma-1)}}(12-\alpha^2)},\label{39}\\
G&=&(1+\alpha^2)\pi^2\Big[\frac{5b^{4\gamma}r_+^{-4\gamma-9}}{16(5+\alpha^2)}+\frac{b^{8\gamma}P\pi r_+^{7-8\gamma}(1-\alpha^2)}{6(7-\alpha^2)}+\frac{3b^{\frac{-2\gamma}{3}}q^{4}(21+\alpha^2)r_+^{\frac{2\gamma-3}{3}}}{2(3+\alpha^2) (12-\alpha^2)}\Big].\nonumber
\\
\label{40}
\end{eqnarray}%

When $p=2,n=7,k=1$, the terms $A$, $B$ and $C$ can be calculated as
\begin{eqnarray}
A&=&\frac{5(1+\alpha^2)^2}{6\pi (1-\alpha^2)},\label{36}\\
B&=&\frac{24-2\alpha^2}{3(\alpha^2-1)},\label{37}\\
C&=&\frac{9q^{4}}{\pi [\frac{2(1+\alpha^2)}{3}]^B b^{\frac{22\alpha^2}{3\alpha^2-3}}(12-\alpha^2)},\label{38}
\end{eqnarray}
with which we can obtain the relevant critical quantities. We list the results for different choices of parameters in Table \ref{tb1}.

\begin{table}[!h]
\tabcolsep 0pt
\caption{Critical quantities for $p=2,n=7,k=1$}
\vspace*{-12pt}
\begin{center}
\def\temptablewidth{0.8\textwidth}
{\rule{\temptablewidth}{1pt}}
\begin{tabular*}{\temptablewidth}{@{\extracolsep{\fill}}cccccc}
$\alpha$ & $q$   & $v_c$   &$T_c$ &$P_c$  &$P_cv_c/T_c$ \\   \hline
  $0.10 \; \; \;$ & $1 \; \; \; $ & $1.00397\; \; \;$       &        $0.467517\; \; \;$   & $0.203997\; \; \;$   & 0.438073 \\
    $0.25\; \; \;$  & $1\; \; \;$    & $1.03861\; \; \;$       &        $0.532952\; \;\; $ & $0.226346\; \;\; $    & 0.441099\\
 $0.50\; \; \;$     &  $1\; \; \;$     & $1.14940\; \; \;$        &        $0.859766\; \; \;$ & $0.338196\; \; \;$    & 0.452128 \\
  $0.75\; \; \;$   & $1\; \; \;$      & $1.28750\; \; \;$        &        $2.15943\; \; \;$ & $0.790496\; \; \;$   & 0.471311\\
           $0.50\; \;\; $   & $2\; \; \;$     & $1.59612\; \; \;$       &        $0.619137\; \;\; $  & $0.175381\; \;\; $   & 0.452128\\
                $0.50\; \;\; $   & $3\; \; \;$     & $1.93409\; \;\; $         &        $0.510946\; \;\; $  & $0.119443\; \;\; $ & 0.452128
       \end{tabular*}
       {\rule{\temptablewidth}{1pt}}
       \end{center}
       \label{tb1}
       \end{table}

\begin{figure*}
\centerline{\subfigure[]{\label{1a}
\includegraphics[width=8cm,height=6cm]{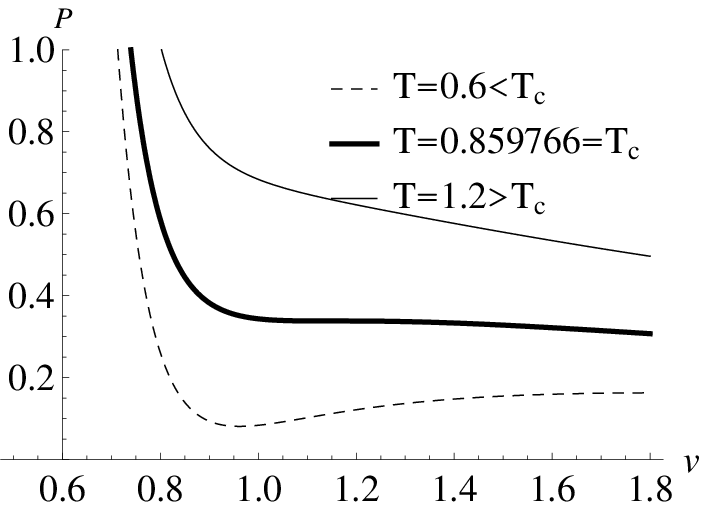}}
\subfigure[]{\label{1b}
\includegraphics[width=8cm,height=6cm]{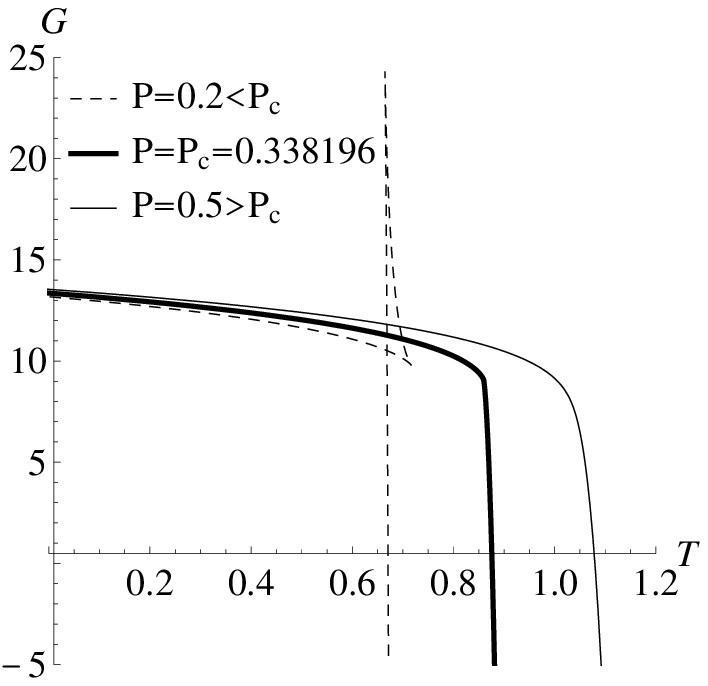}}}
 \caption{(a) $P$ vs. $v$ for $p=2,n=7,q=b=1,\alpha=0.5$ (b) $G$ vs. $T$ for $p=2,n=7,q=b=1,\alpha=0.5$} \label{fg1}
\end{figure*}

$P-v$ graph is plotted in Fig. \ref{1a} while Gibbs free energy graph is depicted in Fig. \ref{1b}. It is shown in $P-v$ graph that the isotherm for the temperature above the critical temperature is monotonically decreasing while the isotherm for the temperature lower than the critical temperature can be divided into three branches. Namely, the stable large radius branch, the stable small radius branch and the unstable medium radius branch. There exists phase transition between the small black hole and the large black hole which is analogous to the van der Waals liquid-gas system. It is shown in Gibbs free energy graph that swallow tail appears when the thermodynamic pressure is lower than the critical pressure.

%%%%%%%%%%%%%%%%%%%%%%%%%%%%%%%%%%%%%%%%%%%%%%%%%%%%%%%%%%%%%%%%%%%%%%%%%%%%%
%%%%%%%%%%%%%%%%%%%%%%%%%%%%%%%%%%%%%%%%%%%%%%%%%%%%%%%%%%%%%%%%%%%%%%%%%%%%%%%%

\section{Critical exponents of topological dilaton black holes with power-law Maxwell field}
\label{Sec4}
The relevant critical exponents are defined as follows
\begin{eqnarray}
C_V&\propto&|t|^{-\alpha},\label{41}\\
\eta&\propto&|t|^{\beta},\label{42}\\
\kappa_T&\propto&|t|^{-\gamma},\label{43}\\
|P-P_c|&\propto&|v-v_c|^{\delta},\label{44}
\end{eqnarray}
where $\alpha$ and $\gamma$ describe the behavior of $C_V$ and $\kappa_T$ respectively while $\beta$ and $\delta$ characterize the behavior of the order parameter $\eta$ and the critical isotherm respectively.

It would not be difficult to understand that the fixed volume specific heat $C_V$ and the critical exponent $\alpha$ both equal to zero since the entropy $S$ does not depend on the Hawking temperature $T$.

To facilitate the calculation of other three critical exponents, one can define
\begin{equation}
t=\frac{T}{T_c}-1,\;\;\epsilon=\frac{v}{v_c}-1,\;\;\hat{p}=\frac{P}{P_c}.\label{45}
\end{equation}
With the above notations, Eq. (\ref{21}) can be expanded as
  \begin{equation}
\hat{p}=1+\hat{p}_{10}t+\hat{p}_{01}\epsilon+\hat{p}_{11}t\epsilon+\hat{p}_{02}\epsilon^2+\hat{p}_{03}\epsilon^3+O(t\epsilon^2,\epsilon^4).\label{46}
\end{equation}
where
\begin{eqnarray}
\hat{p}_{01}&=&\hat{p}_{02}=0,\label{47}\\
\hat{p}_{10}&=&\frac{4\alpha^2+4np-4p-4\alpha^2p}{1+2np-4p+\alpha^2},\label{48}\\
\hat{p}_{11}&=&-\frac{4\alpha^2+4np-4p-4\alpha^2p}{1+2np-4p+\alpha^2},\label{49}\\
\hat{p}_{03}&=&-\frac{4\alpha^2+4np-4p-4\alpha^2p}{1+2np-4p+\alpha^2}+\frac{4(n-2)k(1+\alpha^2)^2}{(n-1)\pi v_c^2P_c(1-\alpha^2)}
\nonumber
\\
&\;&-\frac{2^{p-2}pq^{2p}(n-1)b^{-\frac{2\gamma (np-2p+1)}{(2p-1)(2\gamma-1)}}\Pi}{\pi P_c [\frac{4(1+\alpha^2)}{(n-1)v_c}]^{\frac{2\Pi}{(2p-1)(\alpha^2-1)}}(2p-1)(1-\alpha^2)^2}
\nonumber
\\
&\;&-\frac{(np-3p+p\alpha^2+1)(\alpha^2+2np-4p+1)2^{p}pq^{2p}(n-1)}{24(2p-1)^2(1-\alpha^2)^3\pi b^{\frac{2\gamma (np-2p+1)}{(2p-1)(2\gamma-1)}}[\frac{4(1+\alpha^2)}{(n-1)v_c}]^{\frac{2\Pi}{(2p-1)(\alpha^2-1)}}P_c}.\label{50}
\end{eqnarray}

Since the pressure keeps constant during the phase transition, one can equal the pressure of large black hole with that of small black hole
 \begin{equation}
1+\hat{p}_{10}t+\hat{p}_{11}t\epsilon_l+\hat{p}_{03}\epsilon_l^3=1+\hat{p}_{10}t+\hat{p}_{11}t\epsilon_s+\hat{p}_{03}\epsilon_s^3.\label{51}
\end{equation}

Based on the Maxwell's equal area law, one can obtain
 \begin{equation}
\int^{\epsilon_s}_{\epsilon_l}\epsilon \frac{d\hat{p}}{d\epsilon}d \epsilon=0,\label{52}
\end{equation}
where
\begin{equation}
\frac{d\hat{p}}{d\epsilon}=\hat{p}_{11}t+3\hat{p}_{03}\epsilon^2.\label{53}
\end{equation}
Then one can derive
 \begin{equation}
\hat{p}_{11}t(\epsilon^2_s-\epsilon^2_l)+\frac{3}{2} \hat{p}_{03}(\epsilon^4_s-\epsilon^4_l)=0.\label{54}
\end{equation}
Utilizing Eqs. (\ref{51}) and (\ref{54}), one can obtain
\begin{equation}
\epsilon_l=-\epsilon_s=\sqrt{\frac{-\hat{p}_{11}t}{\hat{p}_{03}}}.\label{55}
\end{equation}
Then
\begin{equation}
\eta=v_l-v_s=v_c(\epsilon_l-\epsilon_s)=2v_c\epsilon_l\propto\sqrt{-t},\label{56}
\end{equation}
According to the definition of $\beta$, one can draw the conclusion that $\beta=1/2$.

Since the isothermal compressibility coefficient
\begin{equation}
\kappa_T=\left.-\frac{1}{v}\frac{\partial v}{\partial P}\right|_{v_c}\propto \left.-\frac{1}{\frac{\partial \hat{p}}{\partial \epsilon}}\right|_{\epsilon=0}=-\frac{1}{\hat{p}_{11}t},\label{57}
\end{equation}
one can conclude that $\gamma=1$.

 Substituting $t=0$ into Eq. (\ref{46}), one can get
\begin{equation}
\hat{p}-1=\hat{p}_{03}\epsilon^3,\label{58}
\end{equation}
According to the definition of $\delta$, one can derive that $\delta=3$.

The above results of critical exponents are totally the same as those of other AdS black holes in former literatures~\cite{Kubiznak,Gunasekaran}, showing the effect of mean field theory.

\section{Conclusions}
\label{Sec5}
    In this paper, we extend the former research~\cite{Sheykhi8} to the extended phase space and probe the effects of power-law Maxwell field on the Van der Waals like phase transition of higher-dimensional dilaton black holes. Treating the cosmological constant as thermodynamic pressure, we derive the explicit expression of its conjugate quantity-thermodynamic volume. It is shown that the thermodynamic volume of dilaton black holes with power-law Maxwell field is exactly the same as that of Einstein-Maxwell-dilaton black holes~\cite{Zhao}. This result implies that power-law Maxwell field does not affect the thermodynamic volume.

The equation of state becomes quite complicated when the dilaton field and Maxwell field are both taken into consideration. So it is a nontrivial task to obtain the analytic expressions of the critical quantities. By introducing some notations to simplify the equation of state, we successfully derive the analytic solutions of critical point. Furthermore, we carry out some check to ensure that these critical quantities are positive. It is shown that the constraint on the parameters turns out to be $0<\alpha^2<1$, which is more tighter than that in the non-extended phase space. It is also shown that these critical quantities are influenced by the power-law Maxwell field while they coincide with those of charged dilaton AdS black holes~\cite{Zhao} when $p=1$. Especially when $k=1,n=3,p=1,\alpha=\gamma=0$, they recover the result of four-dimensional RN-AdS black holes~\cite{Kubiznak}. Moreover, the analytic expression of the ratio $P_cv_c/T_c$ is derived with corrections due to the effect of power-law Maxwell field.

As a specific example, the case $p=2,n=7,k=1$ is discussed thoroughly and the results for different choices of parameters are listed in Table~\ref{tb1}. $P-v$ graph and Gibbs free energy graph are also depicted to provide an intuitive understanding of the possible phase transition. Both the isotherm and Gibbs free energy graph exhibit behaviors analogous to the van der Waals liquid-gas system when the temperature is lower than the critical temperature. Critical exponents are also calculated to probe the critical behavior near the critical point. It is shown that these exponents coincide with those of other AdS black holes, showing the powerful influence of mean field theory.

To summarize, our research discloses the effects of power-law Maxwell field on the Van der Waals like phase transition of higher dimensional dilaton black holes. It will help further understand the rich physics of both the power-law Maxwell field and dilaton field. It will also enhance the understanding of the close relation between AdS black holes and liquid-gas systems.

 \section*{Acknowledgements}
The authors want to express their sincere gratitude to both the editor and the referee for their joint effort to improve the quality of this paper significantly. This research is supported by Guangdong Natural Science Foundation (Grant No.2015A030313789) and Department of Education of Guangdong Province of China(Grant No.2014KQNCX191). It is also supported by \textquotedblleft Thousand Hundred Ten\textquotedblright \,Project of Guangdong Province.

\end{document}